\documentstyle[12pt]{article}
\begin{document}
\def\d{{\rm d}}
\def\e{{\rm e}}
\def\bra#1{\langle #1|}
\def\ket#1{|#1\rangle}
\def\mittel#1{\langle#1\rangle}
\def\r#1{(\ref{#1})}
\def\halb{{\textstyle{1\over 2}}}
\baselineskip=15pt
\renewcommand{\thefootnote}{\fnsymbol{footnote}}
\begin{titlepage}
\begin{flushright}
DUKE-TH-93-55 \\
hep-ph/9402256 \\
February 1994
\end{flushright}
\vspace{1cm}
\begin{center}
{\Huge Pair production in the quantum Boltzmann equation}
\\
\vspace{1cm}
{Jochen Rau\footnote{present address:
Max-Planck-Institut f\"ur Kernphysik,
Postfach 103980, 69029 Heidelberg, Germany}\footnote{e-mail:
rau@eu10.mpi-hd.mpg.de}}
\\
{\sl Physics Department\\
Duke University, Durham, NC 27708--0305}
\end{center}
\vspace{1cm}
\centerline{\bf Abstract}
A source term in the quantum Boltzmann equation, which
accounts for the spontaneous creation of $e^+e^-$-pairs
in external electric fields, is derived from first principles
and evaluated numerically.
Careful analysis of time scales reveals that this source term is
generally non-Markovian.
This implies in particular that there may
be temporary violations of the $H$-theorem.
\\
\vfill\noindent
{\em PACS numbers: 12.38.Mh, 25.75.+r, 52.25.Dg, 05.70.Ln}
\end{titlepage}
The evolution of the quark-gluon plasma, believed to be formed in the
course of relativistic heavy-ion collisions,
is commonly
described by means of a transport equation
\cite{bialas:czyz:1,bialas:czyz:2,kajantie:matsui,gatoff:etal}.
It is well understood how a transport equation can account for
acceleration in
external fields, scattering,
or \mbox{(hadro-)}chemical reactions of the microscopic constituents.
There is, however, another physical process which becomes increasingly
important at high energies:
regions of very large chromoelectric
field strength may develop and subsequently decay
by emitting quark-antiquark pairs
\cite{ehtamo,biro:etal}.
This gives rise
to the fragmentation of chromoelectric flux tubes (`strings'),
a mechanism
frequently invoked to model hadron production
\cite{casher:etal,glendenning:matsui,banerjee:etal,kornel:string}.
How such spontaneous creation of particles
can be incorporated into a transport equation is still not fully
understood.

Clearly, the transport equation has
to be modified by a source term.
What is this source term? How can it be derived from the
underlying microscopic dynamics?
These issues have recently been approached
in a Wigner function formulation
\cite{eisenberg:kaelbermann,rafelski:etal,kluger:1,kluger:2,kluger:central}.
But, aside from the fact that it lacks an intuitive
probabilistic interpretation, this approach
suffers from several practical limitations.
The source term cannot be determined completely:
it is not known how the
longitudinal momenta of the produced particles are distributed.
It has been suggested that the distribution is a $\delta$-function
\cite{casher:etal};
but while such an ansatz
may be useful for practical purposes
\cite{kluger:1,kluger:2,kluger:central},
it is certainly not exact.
Furthermore, an interplay
of pair creation and collisions
--- possibly leading to a modification of the source term ---
has not yet been considered.
And finally, the Wigner description is not suited for discussing
the apparent irreversibility of the particle creation process or the
associated generation of entropy.
Since pair creation in an
external field is merely a single-particle problem (see below),
the Wigner function retains complete information about the
microscopic state of the system.
Yet irreversibility never manifests itself
on the microscopic level;
it only emerges after a suitable coarse-graining.

I choose a different approach.
In collision experiments one usually
measures the momentum distribution
of the outgoing particles; i.e., one determines
the occupation
$n_\mp(\vec p,t):=\mittel{N_\mp(\vec p)}(t)$
of the various momentum
states,
with the number operators given by
\begin{eqnarray}
N_-(\vec p)&:=&\sum_{m_z} a^\dagger(\vec p,m_z) a(\vec p,m_z)
\quad,
\nonumber \\
N_+(\vec p)&:=&\sum_{m_z} b^\dagger(\vec p,m_z) b(\vec p,m_z)
\quad.
\end{eqnarray}
(Here $a$ and $b$ denote particle and antiparticle field operators,
respectively, $\vec p$ the momentum and $m_z$ the spin component.)
This suggests attempting to describe
the evolution of the occupation numbers $\{n_\mp(\vec p,t)\}$
directly
and to derive a kinetic equation for them
--- including the source term --- from first principles.
I will do so with the help of
a very powerful and broadly applicable tool: the
so-called projection method.
This method, pioneered by Nakajima \cite{nakajima:1},
Zwanzig \cite{zwanzig:1,zwanzig:2,zwanzig:4}
and others
\cite{mori:1,robertson:pr1,robertson:pr2,robertson:jmp,kawasaki:gunton,grabert:book},
is based on projecting the motion of the quantum system onto
a low-dimensional subspace (the `level of description')
of the space of observables (Liouville space).
It allows for a clear definition of
crucial concepts like the memory time or the
coarse-grained entropy, making it especially suited
for an investigation of the irreversible features of the dynamics.

As in the works cited above, my
investigation is based on a simple model from
quantum electrodynamics.
I consider
the creation of $e^+e^-$-pairs in a
homogeneous, time-independent electric field $\vec E$,
a process often referred to as the Schwinger mechanism
\cite{sauter:1,heisenberg:euler,schwinger:1}.
The starting point is the
Dirac equation
\begin{equation}
i\hbar{\d\over\d t}\ket{\psi(t)}=H\ket{\psi(t)}
\end{equation}
with
\begin{equation}
H=\hat{\vec p}\cdot\vec\alpha+m\beta+qA_0
\end{equation}
and $A_0(\vec r)=-\vec E\cdot\vec r$
($q=-|e|$ for electrons).
We further define
${\vec p(t)}:={\vec p+q\vec E t}$,
the transverse energy
$\epsilon_\perp:=\sqrt{m^2+p_\perp^2}$,
the total kinetic energy
$\epsilon[\vec p(t)]:=
\sqrt{\epsilon_\perp^2+p_{\|}(t)^2}$,
and the dynamical phase
\begin{equation}
\phi_{fi}:=
-{1\over\hbar}\int_{t_i}^{t_f}\d t'\,
\epsilon[\vec p(t')]
\quad.
\end{equation}
`Longitudinal' and `transverse' refer to the direction of the electric field.

The eigenstates $\ket{i,\pm}\equiv \ket{\vec p(t_i),m_z,\pm}$
which correspond to momentum $\vec p(t_i)$,
spin component $m_z$
and positive or negative energy $\pm\epsilon[\vec p(t_i)]$,
evolve according to
\begin{equation}\label{evol:law}
U(t_f,t_i)
\left(
\begin{array}{c}
\ket{i,+} \\
\ket{i,-}
\end{array}
\right)
=
\left(
\begin{array}{cc}
\alpha_{fi} & \beta_{fi} \\
-\beta^*_{fi} & \alpha^*_{fi}
\end{array}
\right)
\left(
\begin{array}{cc}
\e^{+i\phi_{fi}} & 0 \\
0 & \e^{-i\phi_{fi}}
\end{array}
\right)
\left(
\begin{array}{c}
\ket{f,+} \\
\ket{f,-}
\end{array}
\right)
\quad.
\end{equation}
The evolution thus mixes
positive and negative energy eigenstates,
with respective amplitudes $\alpha_{fi}$ and
$\beta_{fi}$;
$|\beta_{fi}|^2$ equals
the probability for having created an $e^+e^-$-pair
with (final) momenta $\pm\vec p(t_f)$ during the time
interval $[t_i,t_f]$.
The amplitudes are determined by
the differential equation
\begin{equation}\label{diff:equation}
\left(
\begin{array}{c}
\dot\alpha_{fi} \\ \dot\beta_{fi}
\end{array}
\right)
=
{qE\over 2}\cdot
{\epsilon_\perp\over \epsilon[\vec p(t_f)]^2}
\left(
\begin{array}{cc}
0 & -\e^{-i2\phi_{fi}} \\
\e^{+i2\phi_{fi}} & 0
\end{array}
\right)
\left(
\begin{array}{c}
\alpha_{fi} \\ \beta_{fi}
\end{array}
\right)
\end{equation}
with initial conditions
$\alpha_{ii}=1$ and $\beta_{ii}=0$,
and the dot indicating differentiation with respect to $t_f$.

In view of applying the projection method,
the above results have to be translated into
the language of field operators.
To do so, I will use the formulation of
quantum statistical mechanics in Liouville space
\cite{fick:processes}.
There the evolution of (Heisenberg picture)
operators is determined by the so-called super-operators
${\cal L}$ (`Liouvillian') and
${\cal U}$;
these super-operators play a role analogous to that
of $H$ and $U$ in Hilbert space.
Employing the shorthand notations
$a_j\equiv a(\vec p(t_j),m_z)$ and $b_{-j}\equiv
b(-\vec p(t_j), -m_z)$ for the particle and antiparticle field operators,
and making use of the general rule
${\cal U}(t_2,t_1)a^\dagger (\psi)=a^\dagger (U(t_2,t_1)\psi)$,
one finds
\begin{equation}\label{bogoliubov:1}
{\cal U}(t_2,t_1)
\left(
\begin{array}{c}
a^\dagger_1 \\
b_{-1}
\end{array}
\right)
=
\left(
\begin{array}{cc}
\alpha_{21} & \beta_{21} \\
-\beta^*_{21} & \alpha^*_{21}
\end{array}
\right)
\left(
\begin{array}{cc}
\e^{+i\phi_{21}} & 0 \\
0 & \e^{-i\phi_{21}}
\end{array}
\right)
\left(
\begin{array}{c}
a^\dagger_2 \\
b_{-2}
\end{array}
\right)
\quad;
\end{equation}
the evolution law for $(a, b^\dagger)$ follows by
Hermitian conjugation.
Thus pair creation can be described by
a time-dependent Bogoliubov transformation
\cite{fetter:book}.

The Liouvillian
\begin{equation}
{\cal L}=i\left.{\partial\over\partial t_2}\right|_{t_2=t_1}
{\cal U}(t_2,t_1)
\end{equation}
may be written as the sum
\begin{equation}
{\cal L}={\cal L}_{\rm diag}+\delta{\cal L}
\end{equation}
of a diagonal part, responsible for acceleration, and an
off-diagonal part which
is responsible for the mixture of particle and antiparticle
states, i.e., for pair creation.
With the definition
$\dot\beta_{11}:=
\left.\dot\beta_{21}\right|_{t_2=t_1}$,
the latter is given by
\begin{equation}
\delta{\cal L}
\left(
\begin{array}{c}
a^\dagger_1 \\
b_{-1}
\end{array}
\right)
=i
\left(
\begin{array}{cc}
0 & \dot\beta_{11} \\
-\dot\beta^*_{11} & 0
\end{array}
\right)
\left(
\begin{array}{c}
a^\dagger_1 \\
b_{-1}
\end{array}
\right)
\quad.
\end{equation}

Starting from the above microscopic equations, we now want to
derive a kinetic equation for the
occupation numbers
$n_\mp(\vec p,t)$.
Provided the initial state is the vacuum,
\begin{equation}
{\rho(t_0)}=\ket 0\bra 0
\quad,
\end{equation}
momentum and charge conservation dictate
$n_+(\vec p,t)=n_-(-\vec p,t)$
for all later times $t$;
it then suffices to consider
the evolution of only, say, the electron occupation
numbers $n_-(\vec p,t)$.
Their evolution equation must have the structure
\begin{equation}
\dot n_-(\vec p,t)
+q\vec E\cdot\nabla_{\vec p}\,n_-(\vec p,t)
=
\dot n_-^{\rm sou}(\vec p,t)
\end{equation}
with some source term
$\dot n_-^{\rm sou}(\vec p,t)$.
Since we know that this source term accounts for transitions
between positive and negative energy eigenstates, it is
tempting to write down a rate equation of the form
\begin{equation}
\dot n^{\rm sou}(\vec p,+\epsilon,t)
=
\halb r(\vec p\,)\cdot\left[
n(\vec p,-\epsilon,t)-n(\vec p,+\epsilon,t)\right]
\quad,
\end{equation}
$r(\vec p\,)$ being the respective transition rate.
If this were correct, the identifications
$n(\vec p,+\epsilon,t)\equiv n_-(\vec p,t)$ and
$n(\vec p,-\epsilon,t)\equiv 2-n_+(-\vec p,t)$
would then lead to
\begin{equation}
\dot n_-^{\rm sou}(\vec p,t)
=
r(\vec p\,)\cdot S(\vec p,t)
\end{equation}
with
\begin{equation}
S(\vec p,t)
:=
1-\halb n_-(\vec p,t)
-\halb n_+(-\vec p,t)
\quad.
\label{factor:s}
\end{equation}
Such a source term, however, can only be correct
in the Markovian limit --- an approximation which is {\em not}
always justified.
Careful investigation
\cite{rau:thesis,rau:muller}
reveals that the above ansatz for the source term has to be
modified:
assuming the quasistationary limit ($t_0\to -\infty$) one finds
\begin{equation}
\dot n_-^{\rm sou}(\vec p,t)
=
\int_0^\infty\d\tau\,R(\vec p,\tau)\cdot S(\vec p-q\vec E\tau,t-\tau)
\quad.
\label{source:ansatz}
\end{equation}
This source term involves an integration over the entire
history of the system,
thus accounting for finite memory effects and rendering
the evolution of the occupation numbers
generally {\em non}-Markovian.

The kernel $R(\vec p,\tau)$ can be obtained with the help of
the projection method
\cite{rau:thesis,rau:muller}.
One key ingredient in the derivation is the introduction of
a super-operator ${\cal Q}$ which projects
onto the {\em irrelevant} degrees of freedom;
in our case,
\begin{equation}
{\cal Q}N_-={\cal Q}N_+=0
\quad,
\end{equation}
whereas other combinations of field operators are unaffected:
\begin{equation}
{\cal Q}(a^\dagger b^\dagger)=a^\dagger b^\dagger
\quad,\quad
{\cal Q}(ba)=ba
\quad.
\end{equation}
With this definition one finds
\begin{equation}
R(\vec p,\tau)=
-\bra{0}\delta{\cal L}\exp(i{\cal QLQ}\tau)\delta{\cal L}
N_-(\vec p\,)\ket{0}
\quad,
\label{kernel}
\end{equation}
a general result which
holds for {\em arbitrary} field strengths.

The formal expression for the source term can be easily evaluated
in the limit of weak fields.
Provided $E\ll m^2/\hbar q$, then $|\dot\beta|\ll|\dot\phi|$
and hence $\delta{\cal L}$ may be regarded as a small perturbation.
In this case it is legitimate to replace
\begin{equation}
\exp(i{\cal QLQ}\tau)\,\to\,\exp(i{\cal QL}_{\rm diag}{\cal Q}\tau)
\quad,
\end{equation}
leading to
\begin{equation}\label{source}
\dot n_-^{\rm sou}(\vec p,t)=
4\,{\rm Re}\,\int_0^\infty\!\d\tau\,
\dot\beta^*(-\tau,-\tau)\,
\e^{-i2\phi(-\tau,0)}
\,\dot\beta(0,0)\,
S(\vec p-q\vec E\tau,t-\tau)
\end{equation}
(with $\dot\beta(t_2,t_1)\equiv\dot\beta_{21}$
and $\phi(t_2,t_1)\equiv\phi_{21}$).
This source term is consistent with the Schwinger
formula: assuming that
the system is dilute ($S=1$),
using the differential equation \r{diff:equation}
in the limit $\alpha_{fi}\approx 1$,
and employing the Landau-Zener formula
\cite{zener,landau},
one finds that
\begin{equation}
w:=
{1\over h^3}\int\d^3\!p\, \dot n_-^{\rm sou}(\vec p,t)
=
{(qE)^2\over 4\pi^3\hbar^2}\exp\left(-
{\pi m^2\over\hbar qE}\right)
\quad;
\end{equation}
a result which does indeed agree with the leading term
in the Schwinger formula.

The source term \r{source} may be evaluated numerically.
For simplicity I will take the system to be dilute, $S=1$;
the source term then no longer depends on $t$.
It is convenient to introduce
\begin{equation}
a:=\hbar qE/\epsilon_\perp^2
\end{equation}
and to consider, rather than the source term itself, the
dimensionless quantity
\begin{equation}
\eta(a,p_{\|}/\epsilon_\perp):=
{\epsilon_\perp\over qE} \exp\left({\pi\over 2a}\right)\,
\dot n_-^{\rm sou}(\vec p\,)
\quad.
\end{equation}
The pre-factor in its definition has been chosen such that
for $p_{\|}=0$, $\eta$ is of the order one.
As the pre-factor is independent of $p_{\|}$, $\eta$ will
correctly describe the distribution of the longitudinal momenta
of the produced electrons.
One can show that
\begin{equation}
\label{source:explicit}
\eta(a,p_{\|}/\epsilon_\perp)=
\exp\left({\pi\over 2a}\right)
{1\over 2[1+(p_{\|}/\epsilon_\perp)^2]}
\int_{-\infty}^0 \d x
{1\over\cosh^3\varphi(x,p_{\|}/\epsilon_\perp)}
\cos\left({1\over a}x\right)
\end{equation}
with $\varphi$ defined implicitly as the solution of the equation
\begin{equation}
\sinh\varphi\cosh\varphi+\varphi=
x+(p_{\|}/\epsilon_\perp)\sqrt{1+(p_{\|}/\epsilon_\perp)^2}
+\sinh^{-1}(p_{\|}/\epsilon_\perp)
\quad.
\end{equation}

I calculated $\eta$ numerically, using a combination of
Filon's integration formula \cite{handbook} with an efficient
root-finding algorithm.
The results for weak fields ($a<1$; figure 1)
may at first seem surprising.
Clearly the momentum distribution of the produced electrons
is {\em not} narrowly peaked around $p_{\|}=0$;
it is neither a $\delta$-function nor a thermal distribution.
Rather, electrons are being produced predominantly in the direction of
the external field ($p_{\|}>0$).
Electrons moving in the opposite direction ($p_{\|}<0$) are
being annihilated: for them, the production rate is negative.
Of course, such negative production rates are sensible only if
there are electrons available for annihilation.
In the quasistationary limit this is the case:
electrons which have been emitted with positive momentum
are subsequently being decelerated and may then,
as soon as $p_{\|}<0$,
be (partly) annihilated again;
there remains a small surplus which manifests itself as a positive
total production rate.
As another surprising feature, $\eta$ displays (approximately) periodic
oscillations whose period scales with $a$.
This may be
understood qualitatively if one views
pair creation as a tunnelling process
from the negative to the positive energy continuum \cite{casher:etal}.
The barrier between these continua has a spatial width
of the order $\epsilon_\perp/qE$, inducing a `momentum quantisation'
$\Delta p_{\|}\sim\hbar qE/\epsilon_\perp$ and thus
$\Delta(p_{\|}/\epsilon_\perp)\sim a$.
Interference of multiply reflected electron wave functions then leads
to the observed oscillations.
For a strong field ($a>1$; figure 2),
the naive tunnelling picture breaks down;
both the oscillations and the annihilation of particles (negative rates)
become less pronounced.

As we discussed previously, the source term is generally
non-Markovian. It exhibits two characteristic time scales:
(i) the memory time $\tau_{\rm mem}(\vec p)$,
which corresponds to the temporal extent of each
individual creation process
and which indicates
how far back into the past one has to reach
in order to predict future occupation numbers;
and (ii) the production interval $\tau_{\rm prod}(\vec p)$
--- the inverse of the production rate ---,
which corresponds to
the average time that elapses
{\em between} creation processes and thus
constitutes the typical time scale on which the occupation numbers change.
Only if $\tau_{\rm mem}\ll\tau_{\rm prod}$
can memory effects be neglected and the evolution be
considered approximately Markovian.

In the weak field limit
both time scales can be extracted from the source term \r{source}.
First the {\em memory time:}
The factor
\begin{equation}
\dot\beta^*(-\tau,-\tau)\propto
{(\epsilon_\perp/qE)\over (\tau-p_{\|}/qE)^2+(\epsilon_\perp/qE)^2}
\end{equation}
constitutes a Lorentz distribution
in $\tau$, centered around $p_{\|}/qE$ with width $\epsilon_\perp/qE$.
Significant contributions to the source term thus come from
times $\tau$ which are smaller than $(p_{\|}+\epsilon_\perp)/qE$.
As the typical momentum scale is set by $\Delta p_{\|}\sim\hbar qE/
\epsilon_\perp$, we may conclude
\begin{equation}
\tau_{\rm mem}\sim
{\hbar\over\epsilon_\perp}+{\epsilon_\perp\over qE}
\quad.
\end{equation}
The memory time combines two time scales of different origin.
(i) The time $\hbar/\epsilon_\perp$ is proportional to $\hbar$ and
therefore of quantum mechanical origin.
It corresponds (via the time-energy uncertainty relation)
to the time needed to create a {\em virtual} particle-antiparticle
pair,
and may thus be regarded as the `time between two
production attempts.'
(ii) The time $\epsilon_\perp/qE$, on the other hand,
is independent of $\hbar$ and therefore classical.
It can be interpreted in various
ways, depending on the picture employed to visualize
the pair creation process.
If pair creation is viewed as a
tunnelling process,
the classical memory time coincides with the
time needed for the wave function to
traverse the barrier with the speed of light \cite{olkhovsky}.
Alternatively, pair creation may be viewed as
a non-adiabatic transition between the two time-dependent
energy levels $\pm\epsilon[\vec p(t)]$.
In that case the classical memory time corresponds to
the width of the transition region, i.e., the region of
closest approach of the two levels.
Finding the {\em production interval} is less straightforward.
Assuming $p_{\|}=0$ for simplicity,
again invoking the weak field limit, and exploiting the
fact that the source term must be consistent with the
Schwinger formula, one can show that
\begin{equation}
\label{tauprod}
\tau_{\rm prod}(0,\vec p_\perp)\sim
{\epsilon_\perp\over qE}
\exp\left({\pi\epsilon_\perp^2\over 2\hbar qE}\right)
\quad.
\end{equation}

As long as $E\ll m^2/\hbar q\le
\epsilon_\perp^2/\hbar q$,
the particle creation process is Markovian:
$\tau_{\rm mem}\ll\tau_{\rm prod}$.
In the weak field limit, therefore, the entropy
associated with the coarse-grained level of description
spanned by $\{N_\pm(\vec p)\}$,
\begin{eqnarray}
S_{\rm c.g.}(t)
&:=&
-2k{\Omega\over h^3}\int\d^3\!p\,\left[
{n_-(\vec p,t)\over 2}\ln {n_-(\vec p,t)\over 2}+
\left(1-{n_-(\vec p,t)\over 2}\right)
\ln \left(1-{n_-(\vec p,t)\over 2}\right)\right]
+
\nonumber \\
&&
\,+\,(n_-\leftrightarrow n_+)
\quad,
\end{eqnarray}
obeys an $H$-theorem.
($\Omega$ denotes the volume.)
The monotonous increase of the coarse-grained entropy
explains why spontaneous pair creation is perceived as irreversible.
This apparent irreversibility is, of course, a consequence
of the coarse-graining:
information is being transferred from accessible (slow) to
inaccessible (fast) degrees of freedom.
The slow degrees of freedom
are the occupation
numbers of the various momentum states.
{}From these, information gradually `leaks' into unobserved
degrees of freedom: correlations and rapidly oscillating phases
which entangle the respective wave functions of the members
of a particle-antiparticle pair.

As soon as $E\ge m^2/\hbar q$, the situation changes.
The source term \r{source}, which was derived in the
weak field limit, is then only a rough estimate.
Already this weak-field estimate becomes non-Markovian:
at $E=m^2/\hbar q$
the production interval and the memory time
are of the same order $\tau\sim\hbar/m$.
This is a clear indication that at this point conventional Markovian
transport theories must break down.
There may be temporary violations of the $H$-theorem:
the coarse-grained entropy, while still increasing on average,
may now oscillate (on the same scale $\tau\sim\hbar/m$).
Such oscillations have indeed been observed in numerical
simulations \cite{kluger:central}.
A systematic study of these memory effects
should proceed from the general equations \r{source:ansatz}
and \r{kernel}.
Although such an enterprise is beyond the
scope of this letter, we can already say that
(i) memory effects become significant at large field strengths;
and (ii) the projection method can account for these memory effects
and thus appears to be a suitable tool for their investigation.

The above analysis can be extended to include binary collisions
of the produced particles.
This is done by replacing
${\delta{\cal L}}\to{\delta{\cal L}}+{\cal V}$,
where ${\cal V}$ contains the two-body interaction.
To lowest order perturbation theory,
pair creation and collisions do not interfere \cite{rau:thesis};
the additional interaction
gives rise to a separate collision term.
Like the source term, this collision term
is generally non-Markovian and
must be subjected to
a time scale analysis, leading again to a criterion for the
validity of the Markovian approximation.
One finds that there are two contributions to the
memory time: the average time needed for a particle to pass
through an interaction range, and the typical
`off-shell' time given by the time-energy uncertainty relation.
For the Markovian approximation to be valid, these have to be
smaller than the average time that elapses between two successive
collisions \cite{rau:thesis,daniel}.

Let me summarize the main conclusions.
(i) The source term in the quantum Boltzmann
equation can be derived in an unambiguous
fashion by employing the projection method.
(ii) To lowest order, the
source term is not altered by the presence of collisions.
(iii) In the weak field limit, $E\ll m^2/\hbar q$, the source term
is given by \r{source} or \r{source:explicit}, respectively.
It is then Markovian, and the coarse-grained entropy increases
monotonically.
As information is continuously being transferred to inaccessible
degrees of freedom, spontaneous pair creation appears irreversible.
(iv) But as soon as $E\ge m^2/\hbar q$, there may be sizeable
memory effects, leading to temporary violations of the $H$-theorem.
Their description is beyond the scope of
conventional Markovian transport theories.
A more suitable starting point appears to be
the projection method, in particular equations \r{source:ansatz}
and \r{kernel}.
\\\\
{\sl Acknowledgements.}
I thank B. M\"uller for very valuable help and advice,
and C. Waigl for assistance with the numerical work.
Financial support by the Studienstiftung des deutschen Volkes
and by the U.S. Department of Energy (grant no. DE-FG05-90ER40592)
is gratefully acknowledged.

\newpage
\noindent
{\large\bf Figure Captions}
\\\\
Figure 1: The re-scaled production rate $\eta$ as a function of
$p_{\|}/\epsilon_\perp$ for weak fields ($a=0.2, 0.3, 0.7$).
\\\\
Figure 2: The re-scaled production rate $\eta$ as a function of
$p_{\|}/\epsilon_\perp$ for a strong field ($a=2.9$).
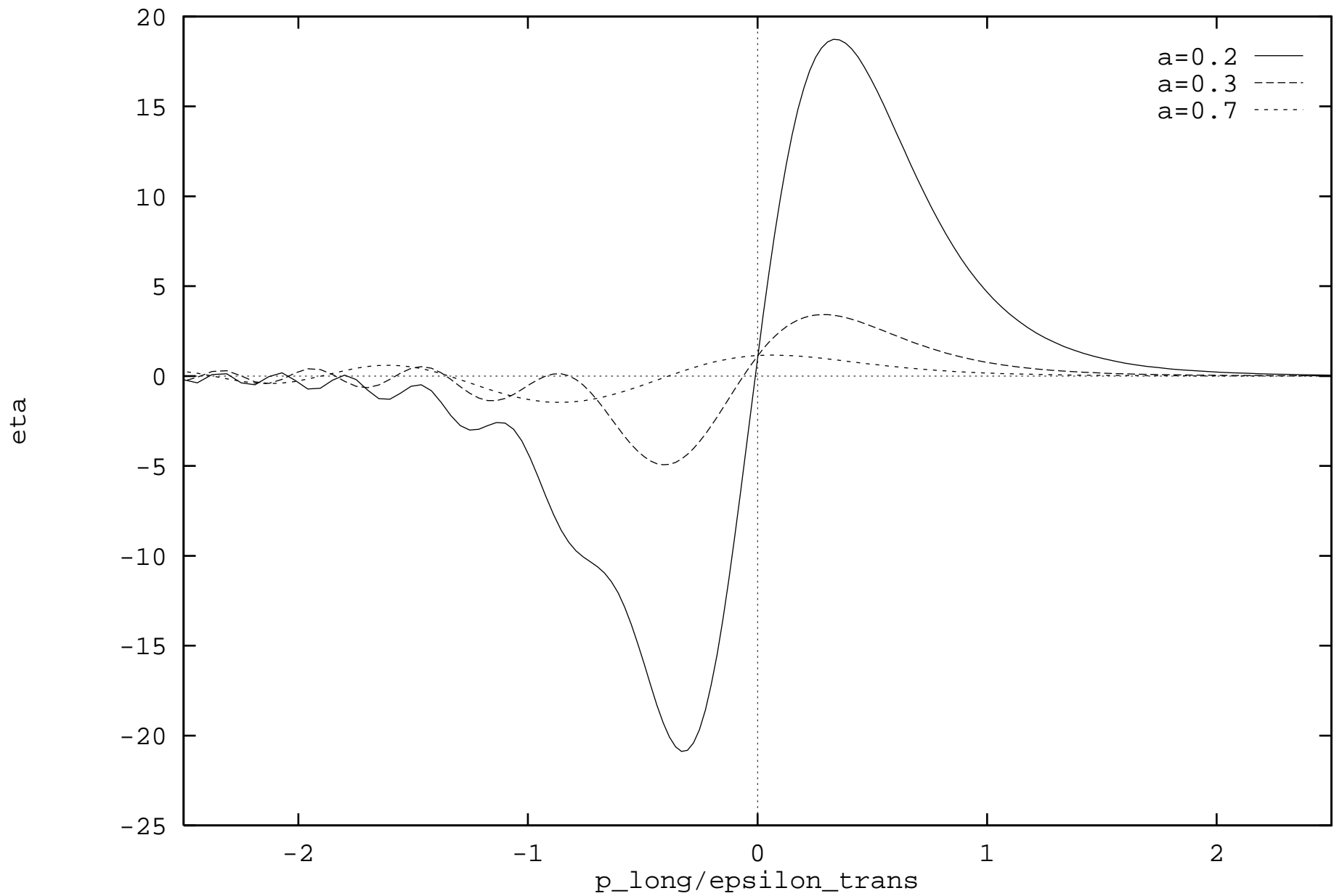
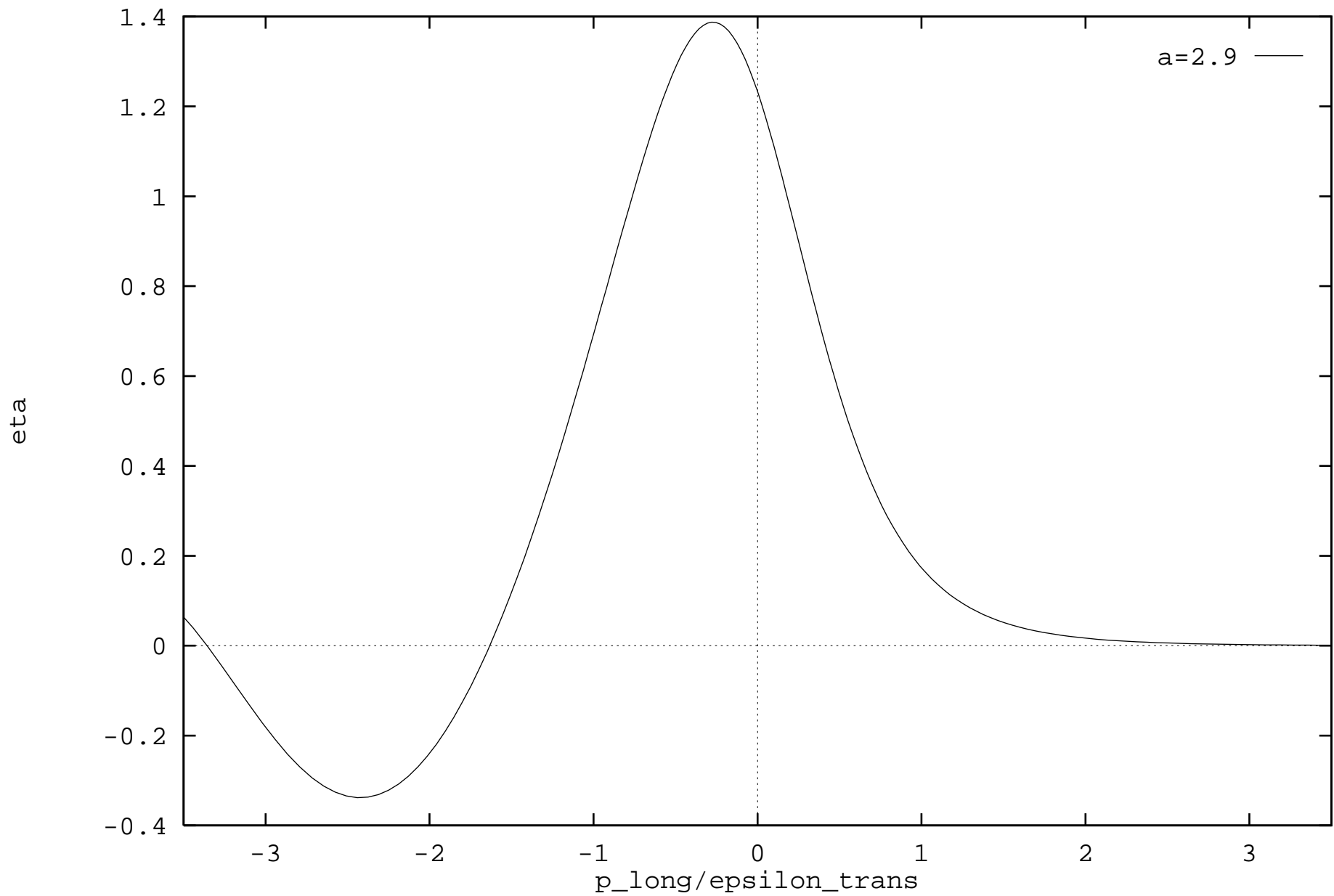
\end{document}